
\input phyzzx

\hsize=6.5truein
\vsize=8.5truein
\hoffset=0.0truecm
\voffset=0.0truecm
\vskip10pt

\chapterstyle={\Roman}

\def\doeack{\foot{Work supported in part by the Department of
Energy under contract DE-AC02-76ERO3130.A021-Task~A}}

\def\spain{\foot{Talk presented at XXII Int. Symposium on Multiparticle
Dynamics, Santiago, Spain, July, 1992, and at Workshop on Small-x and
Diffractive Physics at the Tevatron, FNAL, Batavia, IL, Sept., 1992.}}
\footline={\ifnum\pageno=0\firstfootline\else\otherfootline\fi}
\def\firstfootline{\rm\hss\folio\hss}
\def\otherfootline{\hfil}
\font\tenbf=cmbx10
\font\tenrm=cmr10
\font\tenit=cmti10
\font\elevenbf=cmbx10 scaled\magstep 1
 1
 1
\font\twelverm=cmr10 scaled\magstep 1

\overfullrule=0pt
\def\totalCS{$\sigma_{\scriptstyle {T}}$}

\line{\hfil}

\line{\hfil BROWN HET-889:TA-488,489}
\line{\hfil FERMILAB-CONF-92/391-T}
\line{\hfil hep-ph/9302308}
\vglue 1cm
\parindent=3pc
\baselineskip=10pt
\centerline{\tenbf HETEROTIC POMERON: \hskip10pt A UNIFIED TREATMENT OF}
\centerline{\tenbf  HIGH ENERGY HADRONIC COLLISIONS IN QCD\spain} \vglue 20pt
\centerline{\tenrm GENYA LEVIN} \baselineskip=13pt \centerline{\tenit
Theory Group, Fermi Laboratory, Batavia, IL 60510, USA}
\vglue 10pt\centerline{and}
\vglue 10pt \centerline{\tenrm CHUNG-I TAN\doeack} \baselineskip=13pt
 \centerline{\tenit
Physics Department, Brown University, Providence, RI 02912, USA}

\vskip20pt
\centerline{\tenrm ABSTRACT}
\vglue 0.3cm
{\rightskip=3pc
\leftskip=3pc
\tenrm\baselineskip=12pt
\noindent
A unified treatment of high energy collisions in QCD is presented. Using a
probabilistic approach, we incorporate both perturbative (hard) and
non-perturbative (soft) components in a consistent fashion, leading to a
``Heterotic Pomeron". As a Regge trajectory, it is nonlinear, approaching 1 in
the limit $t\rightarrow -\infty$.\vskip20pt \vglue .6cm} \vskip20pt

\line{\elevenbf 1.  Introduction\hfil} \vglue 0.3cm
\baselineskip=14pt
\twelverm

One of the most striking aspects of high-energy hadron-hadron scattering is
the continued increase of the total cross section \totalCS\ with the energy.
There are currently two seemingly conflicting approaches to high energy
hadronic
collisions in QCD,\REFS\Leningrad{L. V. Gribov, E. M. Levin, and M. G.
Ryskin\journal Phys. Rep. & 100C (83) 1; E. M. Levin and M. G. Ryskin\journal
Phys. Rep. & 189C (90) 267.}\REFSCON\Lipatov{L. N. Lipatov,
review in {\it Perturbative QCD}, ed. A. H. Mueller (World Scientific,
Singapore, 1989), and references therein.}\REFSCON\DPM{A. Capella, U. Sukhetme,
C-I Tan and Tran T. V.\journal Phys. Lett. & B81 (79) 68; for a recent review,
see: {\it Dual Parton Model}, LPTHE-ORSAY-92-38.}\REFSCON\Landshoff{P. V.
Landshoff, Proc. of 3rd Int. Conference on Elastic and Diffractive
Scattering\journal Nucl. Phys. & B12 (90) 397.}\refsend\ as summarized in
Table-I.
  We would like to focus in this paper on
the following  questions: How can qualitative features of rising \totalCS\  be
related to aspects of QCD?  Instead of treating rising \totalCS\ as an isolated
phenomenon, can a  simultaneous description of the elastic and the inelastic
production be achieved by  incorporating   both
perturbative and  nonperturbative aspects of QCD?\Ref\JDBJtwo{ J. D. Bjorken,
{\it Soft and Hard Pomerons: Is There a Distinction?}, Proc. of 4th
International Conference on Elastic and Diffractive Scattering, La Biodola,
Italy, May, 1991.}

It is well understood that the character of  QCD changes depending on  the
nature of available  probes. At short distances, the basic degrees of freedom
are quarks and gluons.  Collisions involving large momentum transfers,
``hard" collisions, can be understood in terms of exchanges of quarks and
gluons. There has  been much recent discussions on the idea of ``semi-hard
processes"  which could account for a large part of the total cross section at
collider energies, with the energy dependence of various cross sections
explained by perturbative QCD motivated calculations.  One usually justifies
this approach by appealing to the work of
  the
``Leningrad" Group.$^{[\Leningrad]}$ In
such a scheme, a rising total cross section is achieved by having a
``hard" Pomeron singularity, the Lipatov Pomeron, above $J=1$.

An equally
compelling argument can also be given in which the dynamical origin of the
increasing total cross sections lies in ``soft" hadronic physics.  As one
moves to larger distance scales, the QCD coupling increases and one  enters the
non-perturbative regime.  The
most promising analytic tool for a non-perturbative treatment of QCD  which
builds in naturally quark-gluon confinement remains the large-$N$ expansion. In
this scheme, model
studies suggest that  the effective degrees of freedom of QCD  can most
profitably be expressed in terms of ``extended objects". Indeed, low-lying
hadron spectrum suggests that they can be understood as ``string excitations".
  In high-energy soft
hadronic collisions\Ref\TAn{C-I Tan, {\it Ideal Gas of Strings and QCD at
Hadronic Scales}, Proc. of Workshop on QCD Vacuum Structure, Paris, France,
June, 1992.} where the interactions are mostly peripheral, it is possible to
``see" the dominant  excitation  in terms of the exchanges of  a soft
Pomeron pole.$^{[\Landshoff]}$  A successful phenomenological model of this
type, particularly for describing production processes, is the Dual Parton
Model
(DPM)$^{[\DPM]}$. A rising total cross section also requires having a soft
Pomeron above $J=1$  in the forward limit, as indicated in the last column in
Table-I.

  \vskip10pt

%
\newbox\hdbox%
\newcount\hdrows%
\newcount\multispancount%
\newcount\ncase%
\newcount\ncols
\newcount\nrows%
\newcount\nspan%
\newcount\ntemp%
\newdimen\hdsize%
\newdimen\newhdsize%
\newdimen\parasize%
\newdimen\spreadwidth%
\newdimen\thicksize%
\newdimen\thinsize%
\newdimen\tablewidth%
\newif\ifcentertables%
\newif\ifendsize%
\newif\iffirstrow%
\newif\iftableinfo%
\newtoks\dbt%
\newtoks\hdtks%
\newtoks\savetks%
\newtoks\tableLETtokens%
\newtoks\tabletokens%
\newtoks\widthspec%
%
%
\immediate\write15{%
CP SMSG GJMSINK TEXTABLE --> TABLE MACROS V. 851121 JOB = \jobname%
}%
%
%
\tableinfotrue%
\catcode`\@=11
%
%
\def\tstrut{\vrule height3.1ex depth1.2ex width0pt}%
\def\and{\char`\&}
\def\tablerule{\noalign{\hrule height\thinsize depth0pt}}%
\thicksize=1.5pt
\thinsize=0.6pt
\def\thickrule{\noalign{\hrule height\thicksize depth0pt}}%
\def\ctr#1{\hfil\ #1\hfil}%
%
%
%
%
\tablewidth=-\maxdimen%
\spreadwidth=-\maxdimen%
\def\tabskipglue{0pt plus 1fil minus 1fil}%
%
%
\centertablestrue%
%
%
%
%
\parasize=4in%
\long\def\para#1{
   {%
      \vtop{%
         \hsize=\parasize%
         \baselineskip14pt%
         \lineskip1pt%
         \lineskiplimit1pt%
         \noindent #1%
         \vrule width0pt depth6pt%
      }%
   }%
}%
\gdef\ARGS{########}
\gdef\headerARGS{####}
\def\@mpersand{&}
{\catcode`\|=13
\gdef\letbarzero{\let|0}
\gdef\letbartab{\def|{&&}}%
\gdef\letvbbar{\let\vb|}%
}
{\catcode`\&=4
\def\ampskip{&\omit\hfil&}
\catcode`\&=13
\let&0
\xdef\letampskip{\def&{\ampskip}}%
\gdef\letnovbamp{\let\novb&\let\tab&}
}
\def\begintable{
   \begingroup%
   \catcode`\|=13\letbartab\letvbbar%
   \catcode`\&=13\letampskip\letnovbamp%
   \def\multispan##1{
      \omit \mscount##1%
      \multiply\mscount\tw@\advance\mscount\m@ne%
      \loop\ifnum\mscount>\@ne \sp@n\repeat%
   }
   \def\|{%
      &\omit\widevline&%
   }%
   \ruledtable
}
\long\def\ruledtable#1\endtable{%
%
%
%
   \offinterlineskip
   \tabskip 0pt
   \def\widevline{\vrule width\thicksize}
   \def\endrow{\@mpersand\omit\hfil\crnorm\@mpersand}%
   \def\crthick{\@mpersand\crnorm\thickrule\@mpersand}%
   \def\crthickneg##1{\@mpersand\crnorm\thickrule
          \noalign{{\skip0=##1\vskip-\skip0}}\@mpersand}%
   \def\crnorule{\@mpersand\crnorm\@mpersand}%
   \def\crnoruleneg##1{\@mpersand\crnorm
          \noalign{{\skip0=##1\vskip-\skip0}}\@mpersand}%
   \let\nr=\crnorule
   \def\endtable{\@mpersand\crnorm\thickrule}%
   \let\crnorm=\cr
%
%
   \edef\cr{\@mpersand\crnorm\tablerule\@mpersand}%
   \def\crneg##1{\@mpersand\crnorm\tablerule
          \noalign{{\skip0=##1\vskip-\skip0}}\@mpersand}%
   \let\ctneg=\crthickneg
   \let\nrneg=\crnoruleneg
   \the\tableLETtokens
%
%
   \tabletokens={&#1}
%
%
   \countROWS\tabletokens\into\nrows%
   \countCOLS\tabletokens\into\ncols%
%
%
   \advance\ncols by -1%
   \divide\ncols by 2%
   \advance\nrows by 1%
%
%
   \iftableinfo %
      \immediate\write16{[Nrows=\the\nrows, Ncols=\the\ncols]}%
   \fi%
%
%
   \ifcentertables
      \ifhmode \par\fi
      \line{
      \hss
   \else %
      \hbox{%
   \fi
      \vbox{%
         \makePREAMBLE{\the\ncols}
         \edef\next{\preamble}
         \let\preamble=\next
         \makeTABLE{\preamble}{\tabletokens}
      }
      \ifcentertables \hss}\else }\fi
   \endgroup
   \tablewidth=-\maxdimen
   \spreadwidth=-\maxdimen
}
\def\makeTABLE#1#2{
   {
   \let\ifmath0
   \let\header0
   \let\multispan0
%
%
   \ncase=0%
   \ifdim\tablewidth>-\maxdimen \ncase=1\fi%
   \ifdim\spreadwidth>-\maxdimen \ncase=2\fi%
   \relax
%
   \ifcase\ncase %
      \widthspec={}%
   \or %
      \widthspec=\expandafter{\expandafter t\expandafter o%
                 \the\tablewidth}%
   \else %
      \widthspec=\expandafter{\expandafter s\expandafter p\expandafter r%
                 \expandafter e\expandafter a\expandafter d%
                 \the\spreadwidth}%
   \fi %
   \xdef\next{
      \halign\the\widthspec{%
      #1
      \noalign{\hrule height\thicksize depth0pt}
      \the#2\endtable
%
      }
   }
   }
   \next
}
\def\makePREAMBLE#1{
   \ncols=#1
   \begingroup
   \let\ARGS=0
   \edef\xtp{\widevline\ARGS\tabskip\tabskipglue%
   &\ctr{\ARGS}\tstrut}
   \advance\ncols by -1
   \loop
      \ifnum\ncols>0 %
      \advance\ncols by -1%
      \edef\xtp{\xtp&\vrule width\thinsize\ARGS&\ctr{\ARGS}}%
   \repeat
   \xdef\preamble{\xtp&\widevline\ARGS\tabskip0pt%
   \crnorm}
   \endgroup
}
\def\countROWS#1\into#2{
   \let\countREGISTER=#2%
   \countREGISTER=0%
   \expandafter\ROWcount\the#1\endcount%
}%
\def\ROWcount{%
   \afterassignment\subROWcount\let\next= %
}%
\def\subROWcount{%
   \ifx\next\endcount %
      \let\next=\relax%
   \else%
      \ncase=0%
      \ifx\next\cr %
         \global\advance\countREGISTER by 1%
         \ncase=0%
      \fi%
      \ifx\next\endrow %
         \global\advance\countREGISTER by 1%
         \ncase=0%
      \fi%
      \ifx\next\crthick %
         \global\advance\countREGISTER by 1%
         \ncase=0%
      \fi%
      \ifx\next\crnorule %
         \global\advance\countREGISTER by 1%
         \ncase=0%
      \fi%
      \ifx\next\crthickneg %
         \global\advance\countREGISTER by 1%
         \ncase=0%
      \fi%
      \ifx\next\crnoruleneg %
         \global\advance\countREGISTER by 1%
         \ncase=0%
      \fi%
      \ifx\next\crneg %
         \global\advance\countREGISTER by 1%
         \ncase=0%
      \fi%
      \ifx\next\header %
         \ncase=1%
      \fi%
      \relax%
      \ifcase\ncase %
         \let\next\ROWcount%
      \or %
         \let\next\argROWskip%
      \else %
      \fi%
   \fi%
   \next%
}
\def\counthdROWS#1\into#2{%
\dvr{10}%
   \let\countREGISTER=#2%
   \countREGISTER=0%
\dvr{11}%
\dvr{13}%
   \expandafter\hdROWcount\the#1\endcount%
\dvr{12}%
}%
\def\hdROWcount{%
   \afterassignment\subhdROWcount\let\next= %
}%
\def\subhdROWcount{%
   \ifx\next\endcount %
      \let\next=\relax%
   \else%
      \ncase=0%
      \ifx\next\cr %
         \global\advance\countREGISTER by 1%
         \ncase=0%
      \fi%
      \ifx\next\endrow %
         \global\advance\countREGISTER by 1%
         \ncase=0%
      \fi%
      \ifx\next\crthick %
         \global\advance\countREGISTER by 1%
         \ncase=0%
      \fi%
      \ifx\next\crnorule %
         \global\advance\countREGISTER by 1%
         \ncase=0%
      \fi%
      \ifx\next\header %
         \ncase=1%
      \fi%
\relax%
      \ifcase\ncase %
         \let\next\hdROWcount%
      \or%
         \let\next\arghdROWskip%
      \else %
      \fi%
   \fi%
   \next%
}%
{\catcode`\|=13\letbartab
\gdef\countCOLS#1\into#2{%
   \let\countREGISTER=#2%
   \global\countREGISTER=0%
   \global\multispancount=0%
   \global\firstrowtrue
   \expandafter\COLcount\the#1\endcount%
   \global\advance\countREGISTER by 3%
   \global\advance\countREGISTER by -\multispancount
}%
\gdef\COLcount{%
   \afterassignment\subCOLcount\let\next= %
}%
{\catcode`\&=13%
\gdef\subCOLcount{%
   \ifx\next\endcount %
      \let\next=\relax%
   \else%
      \ncase=0%
      \iffirstrow
         \ifx\next& %
            \global\advance\countREGISTER by 2%
            \ncase=0%
         \fi%
         \ifx\next\span %
            \global\advance\countREGISTER by 1%
            \ncase=0%
         \fi%
         \ifx\next| %
            \global\advance\countREGISTER by 2%
            \ncase=0%
         \fi
         \ifx\next\|
            \global\advance\countREGISTER by 2%
            \ncase=0%
         \fi
         \ifx\next\multispan
            \ncase=1%
            \global\advance\multispancount by 1%
         \fi
         \ifx\next\header
            \ncase=2%
         \fi
         \ifx\next\cr       \global\firstrowfalse \fi
         \ifx\next\endrow   \global\firstrowfalse \fi
         \ifx\next\crthick  \global\firstrowfalse \fi
         \ifx\next\crnorule \global\firstrowfalse \fi
         \ifx\next\crnoruleneg \global\firstrowfalse \fi
         \ifx\next\crthickneg  \global\firstrowfalse \fi
         \ifx\next\crneg       \global\firstrowfalse \fi
      \fi
\relax
      \ifcase\ncase %
         \let\next\COLcount%
      \or %
         \let\next\spancount%
      \or %
         \let\next\argCOLskip%
      \else %
      \fi %
   \fi%
   \next%
}%
\gdef\argROWskip#1{%
   \let\next\ROWcount \next%
}
\gdef\arghdROWskip#1{%
   \let\next\ROWcount \next%
}
\gdef\argCOLskip#1{%
   \let\next\COLcount \next%
}
}
}
\def\spancount#1{
   \nspan=#1\multiply\nspan by 2\advance\nspan by -1%
   \global\advance \countREGISTER by \nspan
   \let\next\COLcount \next}%
\def\dvr#1{\relax}%
\def\header#1{%
\dvr{1}{\let\cr=\@mpersand%
\hdtks={#1}%
\counthdROWS\hdtks\into\hdrows%
\advance\hdrows by 1%
\ifnum\hdrows=0 \hdrows=1 \fi%
\dvr{5}\makehdPREAMBLE{\the\hdrows}%
\dvr{6}\getHDdimen{#1}%
{\parindent=0pt\hsize=\hdsize{\let\ifmath0%
\xdef\next{\valign{\headerpreamble #1\crnorm}}}\dvr{7}\next\dvr{8}%
}%
}\dvr{2}}
\def\makehdPREAMBLE#1{
\dvr{3}%
\hdrows=#1
{
\let\headerARGS=0%
\let\cr=\crnorm%
\edef\xtp{\vfil\hfil\hbox{\headerARGS}\hfil\vfil}%
\advance\hdrows by -1
\loop
\ifnum\hdrows>0%
\advance\hdrows by -1%
\edef\xtp{\xtp&\vfil\hfil\hbox{\headerARGS}\hfil\vfil}%
\repeat%
\xdef\headerpreamble{\xtp\crcr}%
}
\dvr{4}}
\def\getHDdimen#1{%
\hdsize=0pt%
\getsize#1\cr\end\cr%
}
\def\getsize#1\cr{%
\endsizefalse\savetks={#1}%
\expandafter\lookend\the\savetks\cr%
\relax \ifendsize \let\next\relax \else%
\setbox\hdbox=\hbox{#1}\newhdsize=1.0\wd\hdbox%
\ifdim\newhdsize>\hdsize \hdsize=\newhdsize \fi%
\let\next\getsize \fi%
\next%
}%
\def\lookend{\afterassignment\sublookend\let\looknext= }%
\def\sublookend{\relax%
\ifx\looknext\cr %
\let\looknext\relax \else %
   \relax
   \ifx\looknext\end \global\endsizetrue \fi%
   \let\looknext=\lookend%
    \fi \looknext%
}%
%
%
\def\tablelet#1{%
   \tableLETtokens=\expandafter{\the\tableLETtokens #1}%
}%
\catcode`\@=12

\centerline{Table I. Two Conflicting  Pictures for QCD at High Energies:}
\vskip10pt
\parasize=2.0in
\begintable
\para{Framework:}|\para{Perturbative QCD}|\para{Non-Perturbative QCD}\cr
\para{Emphasis:}|\para {Hard gluons, quarks}|\para{Topological expansion, {\it
etc.}}\cr
\para{Graphs:}|\para{Hard-gluon ladders in LLA}| \para{Soft ladders in
$1/N$ exp.,}\nr
\null|\null| \para{cylinder topology}\cr
\para{Standard Models:}|\para{Leningrad
model$^{[\Leningrad]}$}|\para{Dual Parton Model$^{[\DPM]}$}\cr
\para{Vacuum Exchanges:}| \para{Hard
Pomeron:}| \para{Soft Pomeron: }\nr
\null|\para{$\alpha_L(0)\equiv 1+\Delta_L>1$}| \para{$\alpha_0(0)\equiv
1+\Delta_0>1$}\cr
\para{Total cross sections:}|
\para{Increasing}|\para{Increasing }\cr
\para{Virtuality:}|\para{Increasing $q_{T}^2$}| \para{$q_{T}^2$ fixed and
small}\cr
\para{t-dependence:}|\para{ $\alpha_L(t)\sim $  t-independent}|
\para{$\alpha_0'(0)\sim 0.2 \>{\rm GeV}^{-2}$}\endtable

Rather than treating perturbative and non-perturbative frameworks as
diametrically opposite,  we identify key features of each  which
 allow  a unified treatment of  high energy hadronic collision in QCD.
One of the main puzzles for our current
understanding of high energy hadronic collisions in QCD is the relation, if
any,
between the perturbative (Lipatov) Pomeron and the non-perturbative (soft)
Pomeron.\Ref\CL{J. C. Collins and P. V. Landshoff\journal Phys. Lett. & B276
(92) 196, and references therein. In this interesting work, the emphasis is
on   quantitative questions such as the effects of  cutoffs on the
Lipatov intercept. An attempt  was also made to incorporate the  soft
effect  by isolating it in the ``first rung" of the Lipatov ladder sum.
However, this is insufficient for a systematic unified treatment.  }  This  is
precisely the subject of this talk.  We begin by  briefly reviewing features of
Lipatov and soft Pomerons, both their differences and similarities. We next
show
how the key features of each can be incorporated in a unified treatment leading
to a ``Heterotic Pomeron".
\vglue 0.5cm
\line{\elevenbf 2.  Different Faces of Pomeron in QCD\hfil}
\vglue 0.3cm

 In spite of their apparently different dynamical origins, the hard
and the soft Pomerons share a
structural similarity since they are both generated by summing ladder graphs.
In
Pomeron, one deals with the hard gluon ladders, whereas the soft Pomeron
involves multiperipheral ladders.  In a ladder sum,  one encounters amplitudes
which  at high energy satisfy a  recursion relation  $${\cal A}_n(p,p_0;Q)=\int
d^{4}{p'}K(p,p'; Q){\cal A}_{n-1}(p',p_0; Q), \eqn\ladder$$  where ${\cal
A}_n(p,p_0; Q)$ corresponds to the $n$-rung contribution to the absorptive part
of a non-forward two-to-two amplitude, $(p+Q/2)+(p_0-Q/2)\rightarrow (p-Q/2)+
(p_0+Q/2)$. Let $s\equiv (p+p_0)^2$ and $t\equiv Q^2$, at high energies where
$s>> |t|$, the sum $\sum_n{\cal A}_n(p,p_0;Q)$ can be shown to be power-behaved
in $s$, leading to the respective Pomerons, whose properties are summarized in
Table-I.

Although both pictures can readily lead to increasing total cross
sections,\foot{We emphasize that, with a Pomeron intercept greater than 1,
(either the hard or the soft),  the single-exchange contribution alone would
violate unitarity at sufficiently high energies and screening corrections must
now be taken into account. It is reasonable to assume that this can be carried
out within the context of a Reggeon field theory. If the triple-Pomeron
coupling
is small, a generalized eikonal mechanism becomes operative, leading to an
expanding disk picture for the total cross section. Of course, such a
representation can at best be approximate, possibly meaningful for a limited
range of energies, and it most likely would break down at asymptotic energies.
Our concern here is on the nature of the Pomeron itself and we do not want to
address the question on the precise nature of the screening mechanism. For
simplicity, we shall assume that, at for the current available collider
energies, an eikonal representation is indeed operating.} they
exhibit other distinctive features. Two most important features are: (i) The
production mechanism for a hard Pomeron leads to an increase in ``virtuality",
\ie, the average transverse momentum squared, $\langle q_{T}^2\rangle $
increases with $\log s$, whereas a soft Pomeron exchange corresponds to
processes with limited  $\langle q_{T}^2\rangle $. (ii) The soft Pomeron has a
relatively large slope at $t=0$, whereas the hard Pomeron has a much weaker
$t$-dependence.\foot{ Depending on the approximation used, the hard Lipatov
Pomeron can either be a fixed cut or a series of poles accumulating above
$j=1$.
For our qualitative discussion, we assume that it can be treated as an
effective
$t$-independent $J$-plane singularity. }

These differences could in principle allow one to decide experimentally which
one of these two approaches is  more appropriate  phenomenologically.
Unfortunately, clear cut experimental tests do not exist. In fact, one finds
that it is possible that an additive ``two-component" picture actually work
well
phenomenologically.  However, it should be stressed that a simple additive
approach is in principle  wrong, since these two components must be coupled
through unitarity. It is conceivable that the interference effects between them
would enter only after absorptive corrections have been taken into account, so
that it is meaningful to treat them additively at the level of ``eikonal".

We emphasize that, most significantly, a ``two-gluon ladder" has the
 topology of a ``cylinder" in the color-space, it therefore survives in the
leading large-$N$ limit. Since the soft-Pomeron is also supposed to represent
the effective ladder graphs emerging from the cylinder graphs, it suggests
that,
instead of simply adding, they should be treated as different manifestations of
a more general ``structure" which truly represents the asymptotic behavior of
the cylinder graphs in QCD. We therefore would follow the following strategy
for
unification: \item{(A)} Identify ``distinguishing" features of  each
approach, putting aside quantitative questions such as the precise
value for the ``Lipatov Pomeron intercept",$^{[\CL]}$ \item{(B)} Provide a
``consistent framework" which unifies the key features of each scheme.

\noindent We  carry this out  next  by introducing a  probabilistic
model.

\vglue 0.5cm
\line{\elevenbf 3.  Key Features and Random Walks\hfil}
\vglue 0.3cm

 Let us return to the recursion relation, Eq. \ladder.   At high energies where
$s>> |t|$,
 $p$ and $p_0$ can be decomposed into longitudinal and transverse components.
Whereas the longitudinal components determine  $s$,  their transverse
components,
$\vec q_T$, determine the virtuality of the process. In this limit, $Q$ is also
transverse, \ie, $Q\rightarrow \vec Q_T$ and  $t\sim -\vec Q_T^2$.
Upon taking a
two-dimensional Fourier transform with respect to $\vec Q_T$, \ladder\ can be
written as $${\cal A}_n(s,\vec q_T, \vec b)=\int^s {ds'\over s} \int d^{2}{
q_T'}\int d^{2}{ b'}K(s'/s;\vec q_T, \vec q_T'; \vec b-\vec b'){\cal
A}_{n-1}(s', \vec q'_T;\vec b'). \eqn\laddertwo$$ At high energies, for both
cases, the kernel of this recursion relation
 simplifies which allows  a probabilistic interpretation in
terms of a  ``random walk" picture.

We
begin by first working out the example of a one-dimensional random walk, which
is specified by a normalized elementary one-step probability distribution,
 $\int_{-\infty}^{\infty}\omega(r)dr=1$. The (relative)
probability distribution after $n$ steps in $r$ is then related to that for
$n-1$ steps by a linear relation, $\Psi_n(r)
=\int_{-\infty}^{\infty}dr'\omega(r-r')\Psi_{n-1}(r'). $ We assume
$\omega(r)=\omega(-r)$ so that $\langle
r\rangle_1=\int_{-\infty}^{\infty}r\omega(r)dr=0$.  It follows that $\langle
r\rangle_n=0$   and $\langle r^2\rangle_n$ increases with $n$, the number of
steps taken. Simple  examples for $\omega(x)$ are step-function,
${1\over\lambda}\theta(\lambda-|r|)$, gaussian,
${1\over\sqrt{\pi}\lambda}e^{-r^2/\lambda^2}$, and exponential, ${1\over
2\lambda}e^{-|r|/\lambda}$.
For $n$ large, we can treat  the relative probability as continuous in $n$,
\ie,
$\Psi_n(r)\rightarrow \Psi(n,r)$. Since the  dominant contribution to the
recursive integral    comes from the region where $r\sim r'$, we can expand
the integrand about $(n,r)$, and obtain a diffusion equation
${\partial \Psi(n,r)\over \partial n}=D_r{\partial^2 \Psi(n,r)\over \partial^2
r}$,  where the ``diffusion coefficient" is related to the elementary one-step
fluctuation by  $D_r={1\over 2}\langle r^2\rangle_1$.

A directed random walk corresponds to a situation where  $\omega(y)=0$ for
$y<0$  so that $\lambda\equiv \langle y\rangle_1=\int_{0}^{\infty} y\omega(y)dy
\neq 0$ and  the relative probability after $n$ steps satisfies  the recursion
relation, $\Psi_n(y) =\int_{0}^{y}dy'\omega(y-y')\Psi_{n-1}(y'). $ A simple
consequence of a directed random walk is the fact that the distribution
 in the number of steps taken in reaching  a
large and  fixed distance, $Y$, is Poisson-like. For instance, in the case of
 an exponential step distribution,
$(1/\lambda)e^{-y/\lambda}\theta(y)$, one has $\Psi_{n+1}(y)
=(y^{n}/ \lambda^{n+1} n!)e^{-y/\lambda}$, leading to the result that
the average number of
 steps taken, the average multiplicity, equals to the distance
travelled divided by the
average one-step length, $\langle
n\rangle=y/ \langle
y\rangle_1=y/\lambda.$

Let us next consider a two-dimensional walk in which
$\omega(r,y)\sim \omega_r(r)\omega_y(y)$  where only the walk in the
$y$-direction  is directed, \ie,  $\omega_r(r)=\omega_r(-r)$ and
$\omega_y(y)\propto \theta(y)$.  The joint probability now satisfies the
recursion relation $$\Psi_n(x,y)
=\int_{-\infty}^{\infty}dr'\int_{0}^{y}dy'\omega(r-r', y-y')\Psi_{n-1}(r',y').
\eqn\RWtwo$$ Treating $\Psi_n(r,y)$ as continuous in $n$ for $n$ large  and
expanding the integrand in \RWtwo\ about $(n,r,y)$, one obtains a generalized
diffusion equation  $${\partial \Psi(n,r,y)\over \partial n}=\lambda{\partial
\Psi(n,r,y)\over \partial y}+D_r{\partial^2 \Psi(n,r,y)\over \partial^2 r}.
\eqn\diffusiontwo$$

We  next demonstrate that the structure of both the hard and the soft
Pomerons can be interpreted as simultaneous random walks in appropriate
spaces.
Concentrating first on the hard Pomeron, where Eq. \laddertwo\ corresponds to
the celebrated Lipatov equation. The most remarkable feature of this equation
is
the fact that it is not infrared singular, \ie, the kernel is regular
at $q_T=q_T'$, in spite of the fact that various individual terms contributing
to the kernel  are singular. Much recent discussions have  focused on the
detailed structure of this equation, \eg, the precise intercept of the Lipatov
Pomeron under a variety of physically motivated modifications to this
equation.$^{[\CL]}$ Although these are extremely interesting questions, for our
present purpose, we only  need to identify certain qualitative features  of the
Lipatov equation.

Introducing rapidities
$y=log s$, $y'=log s'$, and $r=log (q_T^2/q_0^2)$, $r'=log
({q'}_T^2/q_0^2)$,\foot{We have introduce here a scale, $q_0$, below which the
LLA used for deriving the Lipatov equation is questionable. We can use this as
a
cutoff below which a non-perturbative description must been used. However, this
will not affect the key ``diffusion" feature which we would like to identify
next.}  the angular part of $\vec q_T$ can be integrated out, and, \laddertwo\
becomes  $${\cal A}_n^{(h)}(y,r, \vec b)\simeq \int_0^y {dy'}
\int_{-\infty}^{\infty} dr'K_L(y-y';r-r'){\cal A}_{n-1}^{(h)}(y',r';\vec b).
\eqn\ladderL$$  The Lipatov kernel is  factorizable,  $K_L(y-y';r-r')\simeq
R_h(y-y')V_h(r-r')$, where $R_h(y)\sim e^y$ and the Fourier transform of
$V_h(r)$, denoted by $\chi(\nu)$,  is analytic for $|{\rm Im} \nu|< 1/2$. The
absence of the $\vec b$-integration, which is an approximation, reflects the
fact
that the Lipatov kernel for \ladder\  asymptotically has a weak $t$-dependence.

Observe that, Eq.
\ladderL\ is structurally similar to \RWtwo. (By dividing an appropriate power
of $s$, $s^{1+\Delta_L(0)}$, it is possible to normalize the kernel $K_L$ so
that  $\int dy\int d r K_L(y,r)=1$. Indeed, one finds that
$\alpha_L\equiv 1+\Delta_L(0)$, $\Delta(0)_L\propto \chi(0)$,  is precisely the
Lipatov Pomeron intercept.) Therefore, the properties of the hard Pomeron can
be
understood in terms of a simultaneous random walk in ``rapidity" and ``log of
virtuality", as summarized in Table-II. In particular, we emphasize that
diffusion in the $r$-space can be  described by an
 equation like \diffusiontwo: $${\partial \Psi(n,\vec b, r,y)\over \partial
n}=\lambda_h{\partial \Psi(n,\vec b, r,y)\over \partial
y}+D_r{\partial^2 \Psi(n,\vec b, r,y)\over \partial^2 r}, \eqn\diffusionL$$
which leads to a spread in virtuality with increasing rapidity:  $\langle
q_T^2\rangle \sim e^{{\rm const.}\sqrt{Y}}$. However, a hard Pomeron does not
lead to diffusion in the impact parameter space.
\vskip10pt
\centerline{Table II. Hard and Soft Pomerons as Random Walks:}
 \vskip10pt
\parasize=1.75in
\begintable

\null | \para{Hard Pomeron} | \para{Soft Pomeron}\cr

\para{Rapidity: $y=\log s$}| \para{Directed-random-walk}
|\para{Directed-random-walk}\cr

\para{Log  of Virtuality:}| \para{Random-walk,} |\para{No,}\nr
\para{$r=log (q_T^2/q_0^2)$}| \para{Diffusion.} |\para{$ q_T^2$ fixed and
small.}\cr

\para{Impact parameter: $\vec b$}| \para{No,} |\para{Random-walk,}\nr
\null| \para{$\vec b$ (approx.) fixed.} |\para{Diffusion.}\endtable
\vskip10pt

We will be more sketchy for the case of a soft Pomeron. Note that, by
definition, a soft ladder structure also involves a strong cutoff in $q_T^2$,
and, for simplicity,  we assume that all $q_T^2$'s are of the order $q_0^2$.
The
kernel must also be cutoff in $t$, thus leading at high energies to a recursion
relation   $${\cal A}_n^{(s)}(y,r_0, \vec b)\simeq \int_0^y {dy'} \int d^2b'
K_s(y-y';\vec b-\vec b'){\cal A}_{n-1}^{(s)}(y',r_0;\vec b'), \eqn\ladderS$$
where $r_0\simeq 0$. That is, a soft Pomeron corresponds to a simultaneous
random walk in the rapidity and the impact parameter space, as summarized in
Table-II. Under a factorizable approximation, one has $K_s(y-y';\vec b-\vec
b')\simeq R_s(y-y')I_s(\vec b-\vec b')$, where $R_s(y)\sim e^{cy}$ and
$I_s(\vec B)$ decreases rapidly for  $\vec B^2$ large.  Instead of \diffusionL,
for $n$ large, $${\partial \Psi(n,\vec b, r_0,y)\over \partial
n}=\lambda_s{\partial \Psi(n,\vec b,r_0, y)\over \partial y}+D_b{\nabla^2
\Psi(n,\vec b,r_0, y)}, \eqn\diffusionS$$ where the diffusion coefficient $D_b$
is related to the one-step fluctuation in $b^2$. This diffusion leads to
$\langle b^2\rangle \propto \log s$, which, when translated back into $t$,
corresponds to the well-known shrinkage of the forward peak due to the exchange
of a soft Pomeron.

\vglue 0.5cm \line{\elevenbf 4.  Unification and Heterotic
Pomeron\hfil} \vglue 0.3cm

We are now in the position to construct a model which incorporates both
diffusion in virtuality, \diffusionL, and diffusion in impact parameter,
\diffusionS.  Note that, whereas diffusion in  $r$  can
take place at any fixed value of  $\vec b$, diffusion in impact parameter space
in a soft process can take place only at small virtuality, $r_0\simeq 0$. This
can be realized in a two-channel simultaneous random walk in the
$y-r-{\vec b}$
space. Let us label the allowed channels by ``s" and ``h", (for soft and hard
respectively.) We introduce four elementary one-step distributions,
$K_{i,j}(y,r,\vec b; y',r',\vec b')$,  the relative probability of starting
from
the $j$th channel at $y',r',\vec b'$ and ending in the $i$th channel at
$y,r,\vec b$ after one step.

For an ordinary random walk, $K_{i,j}$ should depend only on the differences
$y-y',r-r',\vec b-\vec b'$. However, our situation is more restricted, \eg, a
soft process can participate only if the virtuality is small. This can be
simulated by assuming  that  $K_{sh}\propto
\delta(r-r_0)\delta(r'-r_0)$, and  $K_{ss}\propto \delta(r-r_0)$.
Similarly, the fact that  very little  diffusion in impact parameter takes
place
in a hard process can be simulated by assuming that $K_{hh}\propto \delta(\vec
b
-\vec b')\propto K_{hs}$.   Lastly, for directed walk in  rapidity, we must
have
$K_{ij}\propto \theta(y-y')$.  That is, the desired $2\times 2$ one-step
probability distributions can be chosen as  $K_{ss}=g_{ss} R_s(y-y')I_s(\vec
b-\vec b')\delta(r-r_0)$,  $K_{sh}=g_{sh} R_s(y-y')I_s(\vec b-\vec
b')\delta(r-r_0)\delta(r'-r_0)$, $K_{hs}=g_{hs} R_h(y-y')\delta(\vec b-\vec
b')V_h(r-r')$, and  $K_{hh}=g_{hh} R_h(y-y')\delta(\vec b-\vec
b')V_h(r-r')$, where  $R_h$ and $R_s$ can be taken from that used in \ladderL\
and \ladderS\ respectively.

Let the relative probabilities of arriving in the $i$th channel after $n$ steps
be $\Psi_{(n;i)}(y,r,\vec b)$, and let $\Psi_{n}(y,r,\vec b)$
 be the two-vector with $\Psi_{(n;s)}(y,r,\vec b)$ and $\Psi_{(n;h)}(y,r,\vec
b)$ as its upper and lower components respectively. It follows that
$$\Psi_{(n)}(y,r,\vec b)=\int_0^yd y'\int d r'\int d^2b \>K(y,r,\vec b;
y',r',\vec b')\Psi_{(n-1)}(y',r',\vec b'),\eqn\recursionQCD$$ with
$\Psi_{(0;i)}\sim G_i\delta(y)\delta(r-r_0)\delta(\vec b)$. We note that,
because of the structure of $\{K_{ij}\}$, one always has $\Psi_{(n;s)}(y,r,\vec
b)\propto \delta(r-r_0).$ That is, the soft interactions take place at small
virtuality only.

It is appropriate at this point to comment that in specifying $\{K_{ij}\}$ we
have  introduced  a set of symmetric ``coupling matrix", $\{g_{ij}\}$. Whereas
$g_{ss}=0(1)$, the other three must be of the order of the QCD running coupling
constant at a large virtuality, \ie, $g_{hs}\propto g_{sh}\propto g_{hh}\propto
0(\alpha_s)$.  Note also that $\{K_{ij}\}$ are relative probabilities, no
longer
normalized to unity as was done earlier. In particular, the choice $R_h(y)\sim
e^{y}$ and $R_s(y)\sim e^{cy}$, $c\sim 1/2$, correspond precisely to the large
energy behaviors for two-gluon and two-meson exchanges, appropriate for the
hard
and the soft processes respectively. Introduce a complex angular momentum $J$
via a Laplace transform, one has $\tilde R_h(J)=1/(J-1)$ and $\tilde
R_s(J)=1/(J-c)$.

Let $\Psi(y,r,\vec b)=\sum_n \Psi_{n}(y,r,\vec b)$ and denote
$\tilde\Psi(J,\nu,t)$ as its multiple-transform, (Laplace in $y$, Fourier in
$r$
and $\vec b$). It follows from the recursion relation that a formal solution
can
be expressed as $\tilde \Psi(J)=(I-\tilde K)^{-1}\Psi_0$.  We
point out that the high energy behavior of $\Psi(y)$ will be controlled by the
``right-most" singularity of $\tilde\Psi(J)$ in the complex $J$ plane, which is
given by the condition ${\rm det}(I-\tilde K)=0$.

If  soft interactions were turned off, the determinantal condition would lead
to the Lipatov Pomeron, $\alpha_L=1+\Delta_L$, which, as mentioned earlier, is
approximately $t$-independent. In fact, the Lipatov Pomeron, without further
refinements, corresponds to a fixed cut.   Conversely, if the hard processes
were turned off,  one would obtain a soft Pomeron, $\alpha_0(t)=1+\Delta_0(t)$,
which has a ``normal" $t$-slope in the forward region. For simplicity, we
assume
that $0\leq \Delta_s\leq \Delta_L=0(\alpha_s)$. In our unified treatment, a new
singularity, to the right of both $\alpha_L$ and $\alpha_0$, emerges. This new
singularity is a {\it simple pole},  which we referred to as the
``Heterotic Pomeron".  In an approximate treatment, the location of the
Heterotic Pomeron, $\alpha_H(t)\equiv 1+\omega^{*}(t)$, can be found as the
solution to the equation
$\sqrt{\omega^{*}(\omega^{*}-\Delta_L)}(\omega^{*}-\Delta_0(t))=g_{sh}^2G(t)$,
where $G(t)$ is positive and peaked at $t=0$. Details of this analysis will be
presented in a regular publication.\Ref\LT{G. Levin and C-I Tan, in
preparation.}

\vglue 0.5cm \line{\elevenbf 5.
Discussion\hfil} \vglue 0.3cm The fact that Heterotic Pomeron is a pole, with a
slope of the order of that for the soft Pomeron,  might come as a surprise to
some. We will have much more to say about this point elsewhere. Here,  we close
by pointing out some potentially important consequences of our unified
treatment
of QCD at high energies.

Since  Heterotic Pomeron  is a pole, a well-defined perturbative Reggeon
calculus can be carried out. Because of the factorization property, it
naturally
leads to diffractive dissociation events, or more generally, it allows the
study
of ``rapidity gap" physics at collider energies.  We also mention that, in this
unified treatment,$^{[\LT]}$ diffusion in virtuality becomes ``limited",
the Heterotic Pomeron coupling to hadron is ``soft-dominated", and truly hard
processes become dominant only in the region where $|t|\geq 0(\log s)$. As a
Regge   trajectory, the Heterotic Pomeron is nonlinear, approaching 1 in the
limit $t\rightarrow -\infty$ and presumably becoming linear in $t$ for $t$
large
and positive.\REFS\MT{Nonlinear trajectories for mesons in large-$N_c$ QCD have
recently been discussed by M. McGuigan and C. B. Thorn \journal Phys. Rev.
Lett. & 69 (92) 1312. In this study, one concentrates on $t$ large and
negative, without incorporating nonperturbative effects.}\REFSCON\VV{ More
recently, H. Verlinde and E. Verlinde have also re-investigated the high-energy
behavior of QCD in the  region $s>>|t|$, ``QCD at High Energies and
Two-Dimensional Field Theory", PUPT-1319, IASSNS-HEP-92/30,
hep-th/9302104.}\refsend

On a more
phenomenological side, we mention that our treatment allows a systematic
expansion of cross sections in terms of ``hard" and ``soft" events, which goes
beyond the simple ``additive" approach. Furthermore, since Heterotic Pomeron
intercept is  greater than one, absorptive corrections must again be taken into
account. It thus provides a new starting point for handling screening
corrections, which could have a profound effect on our understanding of both
the
near forward hadronic collisions and the small-$x$ physics in deep-inelastic
scattering.

     \refout\end